\documentclass[11pt,peereview,draftclsnofoot,onecolumn]{IEEEtran}

\usepackage{cite}
\usepackage{graphicx}
\usepackage{psfrag}
\usepackage{amsmath,amssymb}
\usepackage{algorithmic}
\def\complex{{\mathbb{C}}}

\def\bfZero{\mathbf{0}}

\def\bfe{{\mathbf{e}}}

\def\bfh{{\mathbf{h}}}

\def\bfn{{\mathbf{n}}}

\def\bfnt{{\tilde{\mathbf{n}}}}

\def\bfs{{\mathbf{s}}}

\def\bfst{{\tilde{\mathbf{s}}}}

\def\bfu{{\mathbf{u}}}

\def\bfy{{\mathbf{y}}}

\def\bfA{{\mathbf{A}}}

\def\bfB{{\mathbf{B}}}

\def\bfH{{\mathbf{H}}}

\def\bfI{{\mathbf{I}}}

\def\bfP{{\mathbf{P}}}

\def\bfQ{{\mathbf{Q}}}

\def\bfS{{\mathbf{S}}}

\def\bfU{{\mathbf{U}}}

\def\bfV{{\mathbf{V}}}

\def\bfLambda{{\mathbf{\Lambda}}}

\DeclareMathOperator*{\defeq}{\triangleq}
\def\hr{^{\mathrm H}}
\newcommand{\prob}[1]{\mathrm{P}\left(#1\right)}
\newcommand{\tprob}[1]{\mathrm{P}(#1)}
\newcommand{\diag}{\mathrm{diag}}

\newtheorem{theorem}{Theorem}

\usepackage[printonlyused]{acronym}
\acrodef{MIMO}{multiple input-multiple output}
\acrodef{QAM}{quadrature amplitude modulation}
\acrodef{FE}{full expansion}
\acrodef{SE}{single expansion}
\acrodef{SD}{sphere decoder}
\acrodef{ML}{ma\-ximum likelihood}
\acrodef{MLD}{maximum likelihood detector}
\acrodef{FSD}{fixed-complexity sphere decoder}
\acrodef{DF}{deci\-si\-on-feedback}
\acrodef{ZF}{zero forcing}
\acrodef{MMSE}{minimum mean square error}
\acrodef{SNR}{signal to noise ratio}
\acrodef{FPGA}{field-programmable gate array}
\acrodef{CDF}{cumulative distribution function}
\acrodef{PSD}{positive semi-definite}
\acrodef{V-BLAST}{vertical-Bell Labs layered space time}

\begin{document}
\sloppy
%
\title{On the Maximal Diversity Order of Spatial Multiplexing with Transmit Antenna Selection}

\author{\authorblockN{Joakim Jald\'{e}n and Bj\"{o}rn Ottersten}\\
\authorblockA{Signal Processing Lab, School of Electrical Engineering,\\
KTH, Royal Institute of Technology,\\
Stockholm, Sweden \\
{\tt [joakim.jalden,bjorn.ottersten]@ee.kth.se}}
\thanks{Manuscript submitted to the IEEE Transactions on Information Theory on December 19, 2006.}}

\markboth{SUBMITTED TO THE IEEE TRANSACTIONS ON INFORMATION THEORY}{}
\maketitle

\begin{abstract}
Zhang et.~al.~recently derived upper and lower bounds on the achievable diversity of an $N_R \times N_T$ i.i.d.~Rayleigh fading multiple antenna system using transmit antenna selection, spatial multiplexing and a linear receiver structure. For the case of $L = 2$ transmitting (out of $N_T$ available) antennas the bounds are tight and therefore specify the maximal diversity order. For the general case with $L \leq \min(N_R,N_T)$ transmitting antennas it was conjectured that the maximal diversity is $(N_T-L+1)(N_R-L+1)$ which coincides with the lower bound. Herein, we prove this conjecture for the zero forcing and zero forcing decision feedback (with optimal detection ordering) receiver structures.
\end{abstract}


\pagebreak
\section{Introduction}

The multiple antennas in a \ac{MIMO} wireless system can be used either to increase the data rate or reliability (diversity) of the wireless link~\cite{TV:05}. In order to capitalize on the benefits offered by the \ac{MIMO} wireless link while maintaining manageable complexity and cost the use of antenna selection has been previously suggested~\cite{MW:04}. In a system using antenna selection only a small subset of the available antennas would typically be used, thereby limiting the number of RF chains required.

In~\cite{ZDZ:06} Zhang et.~al.~rigorously analyzed the maximal achievable diversity for a system transmitting $L$ independent data-streams from $L$ out of $N_T$ possible transmit antennas in conjuncture with linear (decision feedback) processing at the receiver. In particular, for the case of a block i.i.d.~Rayleigh fading channel it was shown that the maximal diversity of such a system is bounded between $M_L \defeq (N_T-L+1)(N_R-L+1)$ and $M_U \defeq (N_T-L+1)(N_R-1)$ where $N_R$ is the number of antennas at the receiver. Since $M_L = M_U$ for $L = 2$ these bounds uniquely determine the maximal diversity in the case of two transmitting antennas and thereby analytically prove some previous observations made in the literature \cite{HSP:01,HP:01}. Further, for the general case where $2 < L < \min(N_R,N_T)$ it was in~\cite{ZDZ:06} conjectured that the maximal diversity conicides with the lower bound, $M_L$. Herein, we extend the anlysis of~\cite{ZDZ:06} by proving this conjecture for the case of the~\ac{ZF} and \ac{ZF}-\ac{DF} receivers (with optimal detection ordering). It should however be noted that the cases of the \ac{MMSE} and \ac{MMSE}-\ac{DF} receivers (although with a fixed detection ordering) also follow from our result by applying the analysis in~\cite{ZDZ:06}.

The structure of this correspondence is as follows. The system model considered is covered in Section~\ref{sec:problem}, mainly in order to the introduce notation. The reader is referred to~\cite{ZDZ:06} for details regarding the systems model and a proper motivation of the problem considered. Our main contribution is then given in Section~\ref{sec:proof} in the form of Theorem~\ref{th:main}.

\section{Problem Statement} \label{sec:problem}

\subsection{System Model}

The case of an $N_R$ by $N_T$ frequency nonselective block Rayleigh fading channel is considered. The channel matrix is denoted $\bfH \defeq \begin{bmatrix}
  \bfh_1 & \bfh_2 & \cdots & \bfh_{N_T} \\
\end{bmatrix}\in \complex^{N_R \times N_T}$ and is assumed constant over a block of $T$ channel uses. Further, the elements of $\bfH$ are modeled as i.i.d.~circularly symmetric complex Gaussian with zero mean and unit variance. The transmitter selects $L \leq \min(N_T,N_R)$ antennas (corresponding to columns of $\bfH$) and transmits independently coded data streams from each antenna. As in~\cite{ZDZ:06}, let $U_j$ denote the $j$th possible antenna subset where
\begin{align}
U_1 = & \{ \bfh_1, \bfh_2, \ldots, \bfh_L \} \nonumber \\
U_2 = & \{ \bfh_1, \bfh_2, \ldots, \bfh_{L-1}, \bfh_{L+1} \} \nonumber \\
\vdots & \nonumber \\
U_{N_U} = & \{ \bfh_{N_T-L+1}, \ldots, \bfh_{N_T} \}
\end{align}
and where $N_U = \tbinom{N_T}{L}$ is the total number of such subsets. The channel can then be modeled according to
\begin{equation} \label{eq:model}
\bfy = \sqrt{\frac{\rho_0}{L}} \bfH_j \bfs + \bfn \,.
\end{equation}
where in the above; $\bfH_j \in \complex^{N_R \times L}$ is the channel matrix containing the columns in the selected subset $U_j$; where $\bfy \in \complex^{N_R \times T}$ is the signal block received during $T$ channel uses; where $\bfs \in \complex^{L \times T}$ is the transmitted signal block; and where $\bfn \in \complex^{N_R \times T}$ is the circularly symmetric complex Gaussian noise which is assumed spatially and temporally white and of unit variance.

At the receiver, a \ac{ZF} front-end is used to separate the transmitted data streams according to
\begin{equation}
\bfst = \bfH^\dagger_j \bfy = \sqrt{\frac{\rho_0}{L}} \bfs + \bfnt
\end{equation}
where $\bfH^\dagger_j = (\bfH_j\hr\bfH_j)^{-1} \bfH_j\hr$ is the pseudo-inverse of $\bfH_j$. Since $L \leq N_R$ by assumption it follows that $\bfQ_j \defeq \bfH_j\hr\bfH_j$ is invertivble with probability one. The effective noise, $\bfnt$, is spatially colored with covariance $\bfQ_j^{-1}$ and the effective post-processing \ac{SNR} of the $k$th data stream is given by
\begin{equation} \label{eq:post-SNR}
\rho_{k}^{(j)} = \left( \frac{\rho_0}{L} \right) \Big/ [\bfQ_j^{-1}]_{kk}
\end{equation}
where $1 \leq k \leq L$~\cite{TV:05,ZDZ:06}. A given data stream, $k$, is said to be in \emph{outage} if the post-processing \ac{SNR} drops below a given threshold, $\gamma > 0$ and the diversity order, $d_k^{(j)}$, of this stream is defined according to
\begin{equation}
d_k^{(j)} = \lim_{\rho_0 \rightarrow \infty} \frac{\ln \tprob{\rho_{k}^{(j)} \leq \gamma}}{\ln \rho_0^{-1}} \, .
\end{equation}
Simially, let $\bar{\rho}^{(j)}$ and $\underline{\rho}^{(j)}$ denote the \emph{maximal} and \emph{minimal} post-processing \ac{SNR}s defined according to
$$
\bar{\rho}^{(j)} \defeq \max_{1 \leq k \leq L} \rho_k^{(j)}
\qquad
\text{and} 
\qquad
\underline{\rho}^{(j)} \defeq \min_{1 \leq k \leq L} \rho_k^{(j)} \, .
$$
Note also that $\underline{\rho}^{(j)} \leq \bar{\rho}^{(j)}$ and that $\bar{\rho}^{(j)} \leq \gamma$ imply that all streams are simultaneously in outage. Thus,
\begin{equation} \label{eq:div-upper}
d_k^{(j)} \leq \bar{d}^{(j)} \defeq \limsup_{\rho_0 \rightarrow \infty} \frac{\ln \tprob{\bar{\rho}^{(j)} \leq \gamma}}{\ln \rho_0^{-1}}
\end{equation}
and
\begin{equation} \label{eq:div-lower}
d_k^{(j)} \geq \underline{d}^{(j)} \defeq \liminf_{\rho_0 \rightarrow \infty} \frac{\ln \tprob{\underline{\rho}^{(j)} \leq \gamma}}{\ln \rho_0^{-1}}
\end{equation}
provides upper and lower bounds on the diversity order of the \ac{ZF} receiver. It also provides upper and lower bounds on the \ac{ZF}-\ac{DF} receiver with optimal ordering since if $\bar{\rho}^{(j)} \leq \gamma$ no data can be reliably decoded and the first data stream decoded is likely to be in error, regardeless of the detection ordering policy applied. Similarly, if $\underline{\rho}^{(j)} \geq \gamma$ all streams can be reliably decoded and~\eqref{eq:div-lower} therefore provides lower bounds on the diveristy of the \ac{ZF} and \ac{ZF}-\ac{DF} receivers. The reader is referred to~\cite{ZDZ:06} for additional details.

\subsection{Problem statement}

In general terms, an \emph{antenna selection policy} is characterized by some (measurable) function $\varphi$
$$
\varphi : \complex^{N_T \times N_R} \mapsto \{ 1, 2, \ldots, N_U \}
$$
which selects an antenna subset, $U_j$, based on the channel matrix realization, $\bfH$, according to $j=\varphi(\bfH)$. In~\cite{ZDZ:06} it is shown that there exists an antenna selection policy, $j = \varphi(\bfH)$, for which
$$
\underline{d}^{(j)} = (N_T-L+1)(N_R-L+1) \, .
$$
This bound in also shown to be tight in the case where $L = 2$ using a geometrical approach. Further, the bound is conjectured to be tight when $L > 2$. Herein, we confirm this conjecture in a positive sense by proving that
$$
\bar{d}^{(j)} \leq (N_T-L+1)(N_R-L+1)
$$
for \emph{any} antenna selection policy, $\varphi$. The proof is given in the following section.

\section{Proof of Conjecture} \label{sec:proof}

In the proof, we let $\succeq$ denote the partial matrix ordering induced by the~\ac{PSD} cone~\cite{BV:04}. For hermitian matrices $\bfA$ and $\bfB$, $\bfA, \bfB \in \complex^{n \times n}$, we write $\bfA \succeq \bfB$ to denote that $\bfA - \bfB$ is~\ac{PSD}. In particular, we will use that $[\bfA]_{kk} \geq [\bfB]_{kk}$ whenever $\bfA \succeq \bfB$ and where $[\bfA]_{kk}$ and $[\bfB]_{kk}$ denotes the $k$th diagonal value of $\bfA$ and $\bfB$. Also, $\bfA \succeq \bfB$ is equivalent to $\bfA^{-1} \preceq \bfB^{-1}$ for strictly positive definite matrices $\bfA$ and $\bfB$ and $\bfA \succeq \bfZero$ if and only if all principal sub-matrices of $\bfA$ are \ac{PSD}~\cite{HJ:85}.

We are now ready to state and prove the contribution of this work which is given by Theorem~\ref{th:main} below. Note also that the theorem yields the recently proved~\cite{JZL:05,ZDH:05} statement that detection ordering can not improve the \ac{ZF}-\ac{DF} diversity order as a special case by selecting $L = N_T \leq N_R$. It should also be noted that the proof of Theorem~\ref{th:main} is similar to a recently submitted proof~\cite{JVZ:05} of this statement but that the antenna selection case represents a non-trivial extension.

\begin{theorem} \label{th:main}
Given an arbitrary antenna selection policy $j = \varphi(\bfH)$ let $\bar{d}^{(j)}$ be defined as in~\eqref{eq:div-upper}. Then
\begin{equation}
\bar{d}^{(j)} \leq (N_T-L+1)(N_R-L+1) \, .
\end{equation}
\end{theorem}
\vspace{2mm}
\noindent \emph{Proof:} Let $\bfQ \defeq \bfH\hr \bfH$, $\bfQ_j \defeq \bfH_j\hr \bfH_j$ and note that $\bfQ_j$ is an $L \times L$ principal sub-matrix of $\bfQ$. Further, let the eigenvalue decomposition of $\bfQ$ be given by
$$
\bfQ = \bfU\bfLambda\bfU\hr
$$
where $\bfLambda = \diag(\lambda_1,\ldots,\lambda_{N_T})$ are the ordered eigenvalues, $\lambda_1 \geq \ldots \geq \lambda_{N_T}$, of $\bfQ$ and where $\bfU = \begin{bmatrix} \bfu_1 & \ldots & \bfu_{N_T} \\ \end{bmatrix}$ are the corresponding eigenvectors. Since $\bfQ$ is unitarily invariant it can be assumed that $\bfU$ is a Haar matrix and independent of $\bfLambda$~\cite{TV:04}. Let $\bfV \defeq \begin{bmatrix}
  \bfu_1 & \ldots & \bfu_{L-1} \\
\end{bmatrix}$ and note that
$$
\bfQ = \sum_{i = 1}^{N_T} \lambda_i \bfu_i \bfu_i\hr
\preceq \sum_{i = 1}^{L-1} \lambda_1 \bfu_i \bfu_i\hr + \lambda_L \bfI
= \lambda_1 \bfV\bfV\hr + \lambda_L \bfI \, .
$$
Let
$$
\bfS \defeq \tfrac{\lambda_1}{\lambda_L} \bfV\bfV\hr + \bfI \, .
$$
and let $\bfS_j$ be the $L \times L$ principal submatrix of $\bfS$ obtained by selecting the rows and columns corresponding to antenna subset $j$. Note that since $\bfQ \preceq \lambda_L \bfS$ it follows that $\bfQ_j \preceq \lambda_L \bfS_j$ and in particular $\bfQ_j^{-1} \succeq \lambda_L^{-1} \bfS_j^{-1}$ which implies that $[\bfQ_j^{-1}]_{kk} \geq \lambda_L^{-1} [\bfS_j^{-1}]_{kk}$ for $k=1,\ldots,L$.

Let $\bfV_j \in \complex^{L \times (L-1)}$ be the matrix consisting of the $L$ \emph{rows} of $\bfV$ corresponding to antenna subset $j$. Note also that $\bfS_j =  \tfrac{\lambda_1}{\lambda_L} \bfV_j\bfV_j\hr + \bfI$. By the matrix inversion lemma it follows that
\begin{align} \label{eq:S-bound}
\bfS_j^{-1} = & (\tfrac{\lambda_1}{\lambda_L} \bfV_j\bfV_j\hr + \bfI)^{-1} \nonumber \\
= & \bfI - \bfV_j (\tfrac{\lambda_L}{\lambda_1} \bfI + \bfV_j\hr\bfV_j)^{-1} \bfV_j\hr \, .
\end{align}
As $\lambda_1 \geq \lambda_L \geq 0$ it follows that
$$
\tfrac{\lambda_L}{\lambda_1} \bfI + \bfV_j\hr\bfV_j \succeq \bfV_j\hr\bfV_j
$$
and therefore
\begin{equation}  \label{eq:inv1}
(\tfrac{\lambda_1}{\lambda_L} \bfI + \bfV_j\hr\bfV_j)^{-1} \preceq (\bfV_j\hr\bfV_j)^{-1}
\end{equation}
which is equivalent to
\begin{equation} \label{eq:inv2}
-(\tfrac{\lambda_1}{\lambda_L} \bfI + \bfV_j\hr\bfV_j)^{-1} \succeq - (\bfV_j\hr\bfV_j)^{-1}\, .
\end{equation}
Note also that the inverse on the right hand side of~\eqref{eq:inv1} exists with probability one due the unitary invariance of $\bfV$ (the probability that any $L$ rows are linearly dependent is zero). Now, inserting~\eqref{eq:inv2} into~\eqref{eq:S-bound} yields
\begin{equation} \label{eq:proj}
\bfS_j^{-1} \succeq \bfI - \bfV_j(\bfV_j\hr\bfV_j)^{-1}\bfV_j\hr \defeq \bfP_j \, .
\end{equation}
In the above, $\bfP_j$ corresponds to a projection onto the null-space of $\bfV_j\hr$ (which has dimension one since $\bfV_j \in \complex^{L \times L-1}$). Note also that for a fixed $j$ (independent of $\bfH$) the distribution of $\bfV_j$ is invariant to multiplication from the right by $L \times L$ unitary matrices. This follows from the unitary invariance of $\bfU$ (and  $\bfV$). Therefore, the null-space of $\bfV_j\hr$ is unitarily invariant and
$$
\prob{[\bfP_j]_{kk} = 0} = 0
$$
for fixed $j$ and $k$ since $[\bfP_j]_{kk} = \bfe_k\hr \bfP_j \bfe_k$ is the squared length of the projection of the $k$th natural basis vector, $\bfe_k$, onto the null-space of $\bfV_j$ (the probability that $\bfe_k$ is completely orthogonal to the null-space is zero). Since there are a finite number of possible $k$ and $j$ it follows that
\begin{equation} \label{eq:low-prob}
\prob{\exists~{k,j}~~[\bfP_j]_{kk} = 0} = 0 \, .
\end{equation}

From~\eqref{eq:low-prob} it follows that there is some constant, $\kappa > 0$, for which
$$
\prob{\exists~{k,j}~~[\bfP_j]_{kk} < \kappa} < 1
$$
or equivalently for which
\begin{equation} \label{eq:fix-bound}
\prob{\forall~{k,j}~~[\bfP_j]_{kk} \geq \kappa} > 0 \, .
\end{equation}
In particular, for $j = \varphi(\bfH)$, it follows that
$$
\prob{[\bfP_j]_{kk} \geq \kappa, ~k=1,\ldots,L} > 0
$$
which states that the probability that all diagonal values of $\bfP_j$ are simultaneously large (in the sense that they are bounded away from zero) is strictly positive.

For notational conveniens in the following, let
$$
\tau \defeq \min_{k,j} [\bfP_j]_{kk} \, .
$$
Since
$$
[\bfQ_j^{-1}]_{kk} \geq \lambda_L^{-1} [\bfS_j^{-1}]_{kk} \geq \lambda_L^{-1} [\bfP_j]_{kk} \geq \lambda_L^{-1} \tau
$$
it follows, by~\eqref{eq:post-SNR}, that
$$
\rho_{k}^{(j)} = \left( \frac{\rho_0}{L} \right) \Big/ [\bfQ_j^{-1}]_{kk} \leq \frac{\lambda_L \rho_0}{\tau L}
$$
for $k=1,\ldots,L$. Thus, if $\tau \geq \kappa$ and $\lambda_L \leq \kappa \gamma L \rho_0^{-1}$ it follows that $\rho_{k}^{(j)} \leq \gamma$ for $k = 1,\ldots,L$. This implies
\begin{align*}
\tprob{ \bar{\rho}^{(j)} < \gamma } \leq & \prob{ \lambda_L \leq \kappa \gamma L \rho_0^{-1} \cap \kappa \leq \tau } \\
= & \prob{ \lambda_L \leq \kappa \gamma L \rho_0^{-1}} \prob{ \kappa \leq \tau }
\end{align*}
where the last equality follows by the independence of $\tau$ (which is a function of $\bfU$) and $\lambda_L$. This implies
$$
\frac{\ln \tprob{\bar{\rho}^{(j)} < \gamma }}{\ln \rho_0^{-1}} \leq 
\frac{\ln \prob{ \lambda_L \leq \kappa \gamma L \rho_0^{-1}}}{\ln \rho_0^{-1}} + 
\frac{\ln \prob{ \kappa \leq \tau }}{\ln \rho_0^{-1}}
$$
where
$$
\limsup_{\rho_0 \rightarrow \infty} \frac{\ln \prob{ \kappa \leq \tau }}{\ln \rho_0^{-1}} = 0
$$
due to~\eqref{eq:fix-bound} and since $\prob{\kappa \leq \tau} > 0$ does not depend on $\rho_0$. Thus,
\begin{align*}
\bar{d}^{(j)} \defeq & \limsup_{\rho_0 \rightarrow \infty} \frac{\ln \tprob{\bar{\rho}^{(j)} < \gamma }}{\ln \rho_0^{-1}} \\
\leq & \limsup_{\rho_0 \rightarrow \infty} \frac{\ln \prob{ \lambda_L \leq \kappa \gamma L \rho_0^{-1}}}{\ln \rho_0^{-1}} + 
\limsup_{\rho_0 \rightarrow \infty} \frac{\ln \prob{ \kappa \leq \tau }}{\ln \rho_0^{-1}} \\
= & \limsup_{\rho_0 \rightarrow \infty} \frac{\ln \prob{ \lambda_L \leq \rho_0^{-1}}}{\ln \rho_0^{-1}} \\
= & (N_T-L+1)(N_R-L+1)
\end{align*}
where the last equality follows from~\cite[Equation (17)]{SAA:05} or as a special case of~\cite[Equation (15)]{ZT:03}. This completes the proof and established the assertion made by the theorem. \hfill $\blacksquare$

\section{Conclusions}

We have proved the conjecture of Zhang~et.~al.~in~\cite{ZDZ:06} regarding the diversity order of spatial multiplexing systems with transmit antenna selection. 

\bibliographystyle{IEEEtran}
\bibliography{refs}

\end{document}